\documentclass[%
reprint,
showpacs,preprintnumbers,
 amsmath,amssymb,
 aps,
pra,
dvipdfmx,
]{revtex4-1}

\usepackage[dvipdfmx]{graphicx}
\usepackage{dcolumn}
\usepackage{bm}
\usepackage[dvipdfmx,
colorlinks=true,
linkcolor=blue,
filecolor=blue,
citecolor=blue,
urlcolor=blue
]{hyperref}

\usepackage{comment}
\begin{document}


\title{Standard quantum limit of angular motion of a suspended mirror\\ and homodyne detection of ponderomotively squeezed vacuum field}

\author{Yutaro Enomoto}
 \email{yenomoto@icrr.u-tokyo.ac.jp}
\author{Koji Nagano}%
\author{Seiji Kawamura}

\affiliation{%
 Institute for Cosmic Ray Research (ICRR), University of Tokyo,\\
 5-1-5 Kashiwa-no-ha, Kashiwa, Chiba 277-8582, Japan
}%




\date{\today}

\begin{abstract}
Compared to the quantum noise in the measurement of the translational motion of a suspended mirror using laser light, the quantum noise in the measurement of the angular motion of a suspended mirror has not been investigated intensively despite its potential importance. 
In this article, an expression for the quantum noise in the angular motion measurement is explicitly derived.
The expression indicates that one quadrature of the vacuum field of the first-order Hermite-Gaussian mode of light causes quantum sensing noise and the other causes quantum backaction noise, or in other words the first-order vacuum field is ponderomotively squeezed.
It is also shown that the Gouy phase shift the light acquires between the mirror and the position of detection of the light corresponds to the homodyne angle. Therefore, the quantum backaction noise can be cancelled and the standard quantum limit can be surpassed by choosing the appropriate position of detection analogously to the cancellation of quantum radiation pressure noise by choosing an appropriate homodyne angle.
\end{abstract}

\maketitle


\section{Introduction}
The standard quantum limit (SQL) is a limitation of quantum noise that universally exists in every precise measurement \cite{Braginsky1995}. The SQL is understood as the trade-off between quantum sensing noise and quantum backaction noise \cite{Braginsky1995}, which arises from the Heisenberg Uncertainty Principle. 
Particularly, quantum noise in the interaction between the translational motion of a suspended mirror and the electromagnetic field of the laser has been investigated intensively for the reduction of quantum noise in interferometric gravitational wave detectors \cite{Danilishin2012} and the realization of a macroscopic quantum mechanical state \cite{Chen2013a}. It was proposed that it is possible to surpass the SQL in the translational motion of a suspended mirror by detecting the electromagnetic field reflected from the mirror with homodyne detection to exploit the correlation between quantum sensing noise and quantum backaction noise that is produced by ponderomotive squeezing of the field \cite{Kimble2002}.
However, quantum noise in the interaction between the angular motion of a suspended mirror and the electromagnetic of the laser has not been investigated sufficiently. Because every degree of freedom is equivalent in the context of the realization of a macroscopic quantum mechanical state, to investigate the quantum noise in the measurement of the angular motion of a suspended mirror is as important as the translational motion of a suspended mirror. 
In this article, therefore, an expression for quantum noise in an optical lever for the measurement of the angular displacement of a suspended mirror is derived. This expression is explained in analogy with the quantum noise in the measurement of the translational motion of a suspended mirror.

\section{Review of notations and background knowledge}
\subsection{Hermite-Gaussian modes and paraxial approximation}
\begin{figure}
\centering
\includegraphics[width=8.5cm,clip]{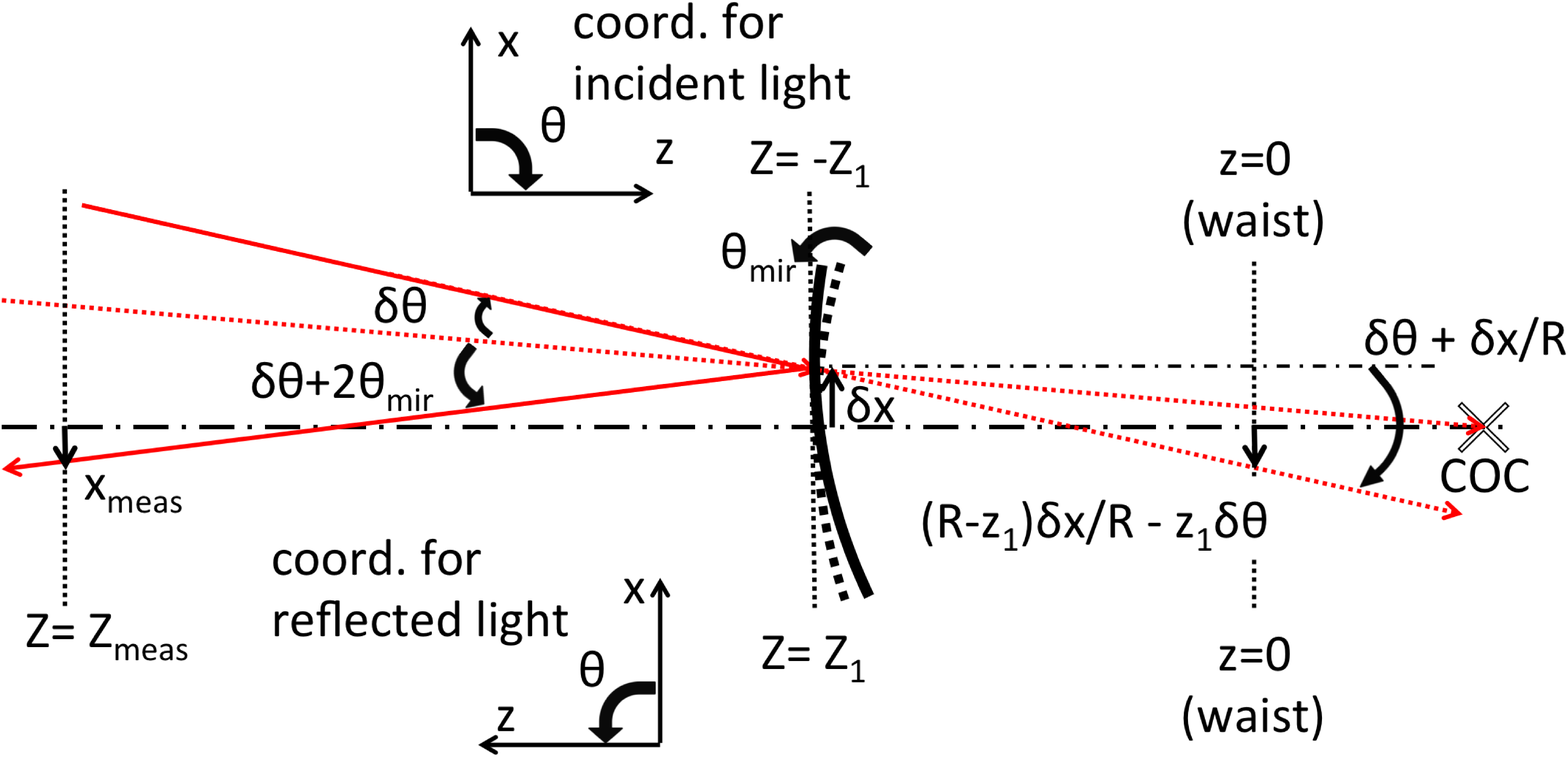}
\caption{
Schematic picture of the configuration of the measurement of the angular motion of a mirror. The transverse displacement and tilt of the incident light measured at the waist and the mirror are shown. The position of the mirror is $z=-Z_1$ in the coordinates of the incident light or $z=Z_1$ in the coordinates of the reflected light. The tilt angle of the mirror is $\theta_\mathrm{mir}$. The reflected light is measured at $z=Z_\mathrm{meas}$. COC represents the center of curvature of the mirror. $R\equiv R(-Z_1)=-R(Z_1)$.}\label{refl2}
\end{figure}

First, let us summarize the background knowledge and the definition of notation. The electric field of a laser is well described with Hermite-Gaussian (HG) modes \cite{Kogelnik1966}. In this article, the following definition of HG modes $U_{lm}$ is used:
\begin{align}
&U_{lm}(x,y,z,t)=u_{lm}(x,y,z)\,\mathrm{e}^{\mathrm{i}\phi_{lm}(x,y,z,t)},\\
&u_{lm}(x,y,z)=\sqrt{\frac{2}{\pi w^2(z)}}\frac{1}{\sqrt{2^ll!2^mm!}}\,\mathrm{e}^{-(x^2+y^2)/w^2(z)}\nonumber\\
&\qquad\qquad\qquad\times H_{lm}\left(\sqrt{2}x/w(z),\sqrt{2}y/w(z)\right),\\
&\phi_{lm}(x,y,z,t)=\omega_0 t-k_0z-k_0\frac{x^2+y^2}{2R(z)}+(l+m+1)\zeta(z).
\end{align}
Here, $\omega_0$ and $k_0$ are angular frequency and wave number of the modes, respectively, and $l$ and $m$ are non-negative integers that characterize modes. The symbols used above are defined as follows:
\begin{align}
\zeta(z)&=\arctan(z/z_0),\\
z_0&=kw_0^2/2,\\
w(z)&=w_0\sqrt{1+z^2/z_0^2},\\
R(z)&=z+z_0^2/z,\\
H_n(x)&=(-1)^n\mathrm{e}^{x^2}\frac{\mathrm{d}^n}{\mathrm{d}x^n}\mathrm{e}^{-x^2},\\
H_{lm}(x,y)&=H_l(x)H_m(y),
\end{align}
where $w_0$ is the beam width at the beam waist ($z=0$). $\zeta,\,z_0,\, w,\, R$, and $H_n$ are the Gouy phase, Rayleigh range, beam width, radius of curvature, and Hermite polynomial of $n$-th order, respectively. $U_{lm}$ is orthonormal in the sense that
\begin{align}
&\int_{-\infty}^\infty\mathrm{d}x\mathrm{d}y\,U^*_{lm}(x,y,z,t)U_{l'm'}(x,y,z,t)\nonumber\\
=&\int_{-\infty}^\infty\mathrm{d}x\mathrm{d}y\,u_{lm}(x,y,z)u_{l'm'}(x,y,z)=\delta_{ll'}\delta_{mm'}\label{ortho}.
\end{align}
An arbitrary electric field $E$ described with HG modes takes the form
\begin{align}
E^{(-)}&=\sqrt{\frac{2\pi I_0}{c}}\sum_{l,m=0}^\infty f^*_{lm}U_{lm}\quad\left(\sum_{l,m=0}^\infty |f_{lm}|^2=1\right), \\
E&=E^{(+)}+E^{(-)}\qquad(E^{(+)}=E^{(-)*}),
\end{align}
where $I_0$ is the power of the light characterized by this $E$ and $f_{lm}$ is a numerical coefficient. Throughout this article, a certain polarization is considered and we deal with the light within the range of paraxial approximation, i.e. an electric field can be regarded as a scalar.
\newline \indent It was shown \cite{Morrison1994} that if we have laser light which can be characterized by the fundamental HG mode and its path is displaced by $\delta x$ and/or tilted by $\delta\theta$ in the $x$-axis direction from the $z$-axis, the electric field of the light takes the form:
\begin{equation}
E^{(-)}\propto\,U_{00}+\left(\frac{\delta x}{w_0}+\mathrm{i}\frac{\delta\theta}{\alpha_0}\right)U_{10}.\label{tilt}
\end{equation}
Here we define $\alpha_0=2/kw_0$. 
Equation (\ref{tilt}) can be derived by expanding $U_{00}(x',y,z',t)$ and taking the terms up to the first order of $\delta x$ and $\delta\theta$ and the leading order of $\alpha_0$, where
\begin{align}
\begin{bmatrix}
z\\
x\\
\end{bmatrix}=
\begin{bmatrix}
\cos\delta\theta&\sin\delta\theta\\
-\sin\delta\theta&\cos\delta\theta\\
\end{bmatrix}
\begin{bmatrix}
z'\\
x'\\
\end{bmatrix}+
\begin{bmatrix}
0\\
\delta x\\
\end{bmatrix}.\label{coord}
\end{align}
Equation (\ref{coord}) gives us an exact interpretation of $\delta x$ and $\delta\theta$; $\delta x$ represents the displacement of the light path from the $z$-axis at the waist of the beam, and $\delta\theta$ is the tilt of the light path from the $z$-axis.
Here and henceforth we assume the displacement and the tilt are small, i.e. $\delta x/w_0\ll1$, $\delta\theta/\alpha_0\ll1$. 
To generalize Eq. (\ref{tilt}), for arbitrary $Z$ let us consider the electric field described by
\begin{align}
&E^{(-)}\propto\,U_{00}+\left(\frac{\delta x}{w(Z)}+\mathrm{i}\frac{\delta\theta}{\alpha(Z)}\right)U'_{10}\label{10d}\\
&=\,U_{00}+\left(\frac{\delta x}{w_0\sqrt{1+(Z/z_0)^2}}+\mathrm{i}\frac{\delta\theta\sqrt{1+(Z/z_0)^2}}{\alpha_0}\right)\nonumber\\
&\qquad\times\left(\frac{1}{\sqrt{1+(Z/z_0)^2}}-\mathrm{i}\frac{Z/z_0}{\sqrt{1+(Z/z_0)^2}}\right) U_{10}\nonumber\\
&=\,U_{00}+\left(\frac{\frac{\delta x}{1+(Z/z_0)^2}+Z\delta\theta}{w_0}+\mathrm{i}\frac{\delta\theta-\frac{(Z/z_0^2)\,\delta x}{1+(Z/z_0)^2}}{\alpha_0}\right)U_{10}\nonumber\\
&=\,U_{00}+\left(\frac{\frac{R(Z)-Z}{R(Z)}\delta x+Z\delta\theta}{w_0}+\mathrm{i}\frac{\delta\theta-\,\delta x/R(Z)}{\alpha_0}\right)U_{10}.\label{10a}
\end{align}
Here we used the relation $\omega_0/\alpha_0=z_0$ and we define:
$\alpha(z)=2/kw(z)$, 
and $U'_{lm}(x,y,z,t;Z)
=U_{lm}(x,y,z,t)\mathrm{e}^{-\mathrm{i}(l+m)\zeta(Z)}$.
Note that all of $U'_{lm}$ have the same phase at $z=Z$. From Eqs. (\ref{tilt}) and (\ref{10a}) the displacement at the beam waist and the tilt of the beam are identified, and the path of the beam described by Eq. (\ref{10d}) is obtained as shown in Fig. \ref{refl2}.
With the help of Fig. \ref{refl2}, the physical interpretation of $\delta x$ and $\delta \theta$ in Eq. (\ref{10d}) is as follows: $\delta x$ represents displacement of the beam at $z=Z$, and $\delta\theta$ represents the tilt angle of the beam measured from the line connecting the center of curvature and the beam position at $z=Z$.

Quantized fields of Hermite-Gaussian modes can be expressed as \cite{Wunsche2004,Aiello2010}
\begin{equation}
\hat{E}_{lm}(x,y,z,t)=\int_0^\infty\frac{\mathrm{d}\omega}{2\pi}\sqrt{\frac{2\pi\hbar\omega}{c}}\left[U^*_{lm}
\hat{a}^{lm}_{\omega}
+\mathrm{H.c.}\right]\label{q}.
\end{equation}
Here $\mathrm{H.c.}$ denotes Hermitian conjugate, and $\hat{a}^{lm}_{\omega}$ and $\hat{a}^{\dagger lm}_\omega$ are the annihilation and creation operators, respectively, which satisfy the commutation relation
\begin{equation}
\left[\hat{a}^{lm}_{\omega},\,\hat{a}^{\dagger l'm'}_{\omega'}\right]=2\pi \delta_{ll'}\delta_{mm'}\delta(\omega-\omega')\label{comm}
\end{equation}
following from the orthonormality relation (\ref{ortho}) of HG modes \cite{Wunsche2004}. Let us define sideband frequency $\Omega$ by $\Omega=\omega-\omega_0$ and let us restrict ourselves to the condition $\Omega\ll\omega_0$, and then we can rewrite Eq. (\ref{q}) as
\begin{align}
\hat{E}_{lm}(x,y,z,t)&=\sqrt{\frac{4\pi\hbar\omega_0}{c}}u_{lm}(x,y,z)\left[\hat{a}^{lm}_1(z,t)\cos\phi_{lm}\right.\nonumber\\
&\left.+\hat{a}_2^{lm}(z,t)\sin\phi_{lm}\right],\label{qq}
\end{align}
where $\hat{a}^{lm}_1(\Omega)=\left(\hat{a}^{lm}_{\omega_0+\Omega}+\hat{a}^{\dagger lm}_{\omega_0-\Omega}\right)/\sqrt{2}$, $\hat{a}^{lm}_2(\Omega)=\left(\hat{a}^{lm}_{\omega_0+\Omega}-\hat{a}^{\dagger lm}_{\omega_0-\Omega}\right)/\mathrm{i}\sqrt{2}$, and then $\hat{a}^{lm}_j(z,t)=\int_0^\infty\frac{\mathrm{d}\Omega}{2\pi} \times \left(\hat{a}^{lm}_j(\Omega)\mathrm{e}^{\mathrm{i}\Omega(z/c-t)}+\mathrm{H.c.}\right)$.
If we decompose $\hat{E}_{lm}$ as $\hat{E}_{lm}=\hat{E}^{(+)}_{lm}+\hat{E}^{(-)}_{lm}$ such that $\hat{E}^{(+)}_{lm}=\hat{E}^{(-)\dagger}_{lm}$, we obtain the following expression from Eq. (\ref{qq}):
\begin{align}
\hat{E}^{(-)}_{lm}(x,y,z,t)&=\sqrt{\frac{\pi\hbar\omega_0}{c}}U_{lm}(x,y,z,t)\nonumber\\
&\times\left[\hat{a}^{lm}_1(z,t)-\mathrm{i}\hat{a}^{lm}_2(z,t)\right].\label{decom}
\end{align}
Following from Eq. (\ref{comm}), the commutation relations
\begin{equation}
\begin{aligned}
\left[\hat{a}^{lm}_1(\Omega),\,\hat{a}^{\dagger l'm'}_2(\Omega')\right]&=\mathrm{i}2\pi\delta_{ll'}\delta_{mm'}\delta(\Omega-\Omega'),\\
\left[\hat{a}^{lm}_1(\Omega),\,\hat{a}^{\dagger l'm'}_1(\Omega')\right]&=\left[\hat{a}^{lm}_2(\Omega),\,\hat{a}^{\dagger l'm'}_2(\Omega')\right]\\
&=2\pi\delta_{ll'}\delta_{mm'}\delta(\Omega-\Omega'),\\
\left[\hat{a}^{lm}_1(\Omega),\,\hat{a}^{l'm'}_2(\Omega')\right]&=0
\end{aligned}\label{com}
\end{equation}
are satisfied.

\subsection{Quantum noise: translation}
Here, in order to clarify the analogy between the quantum noise in the measurement of the translational motion of a mirror and the angular motion of a mirror, the quantum noise in the measurement of the translational motion of a mirror is summarized. The contents summarized here were originally developed by Kimble {\it et al.} \cite{Kimble2002} and we summarize these similarly to Miao \cite{Miao2010}.
Let us consider a model in which the laser light reflected by the mirror is detected at a fixed point in order to determine the position of the mirror. The laser light incident to the mirror is described by the fundamental HG mode. The incident field of the laser light can be expressed as
\begin{equation}
\hat{E}_{\mathrm{in}}(t)=(A+\hat{a}^{00}_1(t))\cos\omega_0t+\hat{a}_2^{00}(t)\sin\omega_0t.
\end{equation}
where $A=\sqrt{2I_0/\hbar\omega_0}$ and $I_0$ is the average power of the light. Here, we omitted the factor $\sqrt{4\pi\hbar\omega_0/c}\,u_{00}$ to ignore spatial dependence of the electric field of the light for simplicity. 
Since the position of the mirror $\Delta \hat{z}_\mathrm{mir}$ can be expressed as 
\begin{equation}
\Delta\hat{z}_\mathrm{mir}(\Omega)=\Delta \hat{z}_\mathrm{ex}-\frac{2\hbar\omega_0A}{cm\Omega^2}\hat{a}^{00}_1(\Omega)
\end{equation}
including the effect of radiation pressure fluctuation of the incident light, the reflected field can be written as
\begin{align}
\hat{E}_{\mathrm{out}}(t)&=(A+\hat{b}^{00}_1(t))\cos\omega_0t+\hat{b}_2^{00}(z,t)\sin\omega_0t,\\
\hat{b}^{00}_1(\Omega)&=\hat{a}^{00}_1(\Omega),\label{pdr1}\\
\hat{b}^{00}_2(\Omega)&=(-\kappa_0\hat{a}^{00}_1(\Omega)+\hat{a}^{00}_2(\Omega))
+\sqrt{2\kappa_0}\frac{\Delta \hat{z}_\mathrm{ex}(\Omega)}{z_\mathrm{SQL}}.\label{pdr2}
\end{align}
Here $m$ is mass of the mirror, 
$\kappa_0=8I_0\omega_0/mc^2\Omega^2$, $z_\mathrm{SQL}=\sqrt{2\hbar/m\Omega^2},$
$\hat{b}^{00}_j(\Omega)$ is the Fourier transform of $\hat{b}^{00}_j(t)\,\,(j=1,2)$, and $\Delta \hat{z}_\mathrm{ex}$ is the displacement caused by an external force.
Then we consider that homodyne detection is conducted for $\hat{b}^{00}_1(\Omega)$ and $\hat{b}^{00}_2(\Omega)$, i.e. the measured value is a linear combination of the two:
\begin{equation}
\hat{b}^{00}_\eta(\Omega)\equiv\hat{b}^{00}_1(\Omega)\cos\eta+\hat{b}^{00}_1(\Omega)\sin\eta,
\end{equation}
where $\eta$ is called the homodyne angle. Homodyne detection is realized by measuring the power of light which is superposition of the quantum field to be measured and local oscillator field with phase difference between these fields controlled at certain value. This phase difference is the homodyne angle. From Eq. (\ref{com}), the one-sided spectral density of $\hat{a}^{lm}_j(t)$ can be expressed as:
\begin{equation}
S^{lm}_{jk}(\Omega)=\delta_{jk}\qquad(j,k=1,2), \label{spectrum}
\end{equation}
and hence the noise spectral density when $\Delta \hat{z}_\mathrm{ex}$ is measured with $\hat{b}^{00}_\eta$ is
\begin{equation}
S_{z,\eta}(\Omega)=\frac{z^2_\mathrm{SQL}}{2\kappa_0}\left[(-\kappa_0+\cot\eta)^2+1\right].\label{00noise}
\end{equation}
If $\eta=\pi/2$, the inequality $S_{z,\eta}(\Omega)\ge z_\mathrm{SQL}^2$ holds for arbitrary $\kappa_0$. 
Note that at the frequency where $\kappa_0=\cot\eta$, radiation pressure noise is cancelled and the SQL, $z_\mathrm{SQL}$, is beaten.

\section{Quantum noise: rotation}
From here, the quantum noise in the measurement of the angular motion of a mirror is discussed.
Let us consider measuring the angular motion of a mirror by detecting the transverse shift of the laser beam reflected by the mirror, as shown in Fig. \ref{refl2}. This is how an optical lever measures the angular motion of a mirror. For simplicity, it is assumed that the radius of curvature of the beam at the mirror and the radius of curvature of the mirror coincide.
First, consider the fluctuation of the transverse displacement and tilt of the incident laser light. In this section we assume the fluctuation is so small that Eqs. (\ref{tilt}) and (\ref{10d}) can be applied. The incident laser light is assumed to be described by the fundamental HG mode and then the electric field of the incident light is written as
\begin{align}
\hat{E}_{\mathrm{in}}(x,y,z,t)&=\sqrt{\frac{4\pi\hbar\omega_0}{c}}\left[u_{00}A\cos\phi_{00}\right.\\
 &\left.+u_{10}\hat{a}^{10}_1(z,t)\cos\phi_{10}+u_{10}\hat{a}_2^{10}(z,t)\sin\phi_{10}\right].\nonumber
\end{align}
Here we omitted the higher order HG modes other than 00 and 10 modes because only these modes are relevant to the transverse displacement and tilt of the laser light in the $x$-axis direction.
If we decompose $\hat{E}_{\mathrm{in}}$ in the same way as Eq. (\ref{decom}) again,
\begin{align}
\hat{E}_{\mathrm{in}}^{(-)}(x,y,z,t)&=\sqrt{\frac{\pi\hbar\omega_0}{c}}\left\{U_{00}(x,y,z,t)A\right.\\
&+U_{10}(x,y,z,t)\left[\hat{a}^{10}_1(z,t)-\mathrm{i}\hat{a}^{10}_2(z,t)\right]\}\nonumber.
\end{align}
In order to know the fluctuation at $z=Z$, defining fundamental-mode-referred quadrature at $z=Z$, $\hat{a}^{lm}_j(z,t;Z)$, by the equation
\begin{align}
&\begin{bmatrix}
\hat{a}^{lm}_1(z,t;Z)\\
\hat{a}^{lm}_2(z,t;Z)\\
\end{bmatrix}\\
&=
\begin{bmatrix}
\cos(l+m)\zeta(Z)&\sin(l+m)\zeta(Z)\\
-\sin(l+m)\zeta(Z)&\cos(l+m)\zeta(Z)\\
\end{bmatrix}
\begin{bmatrix}
\hat{a}^{lm}_1(z,t)\\
\hat{a}^{lm}_2(z,t)\\
\end{bmatrix}\nonumber,
\end{align}
we can express $\hat{E}_{\mathrm{in}}^{(-)}$ as
\begin{align}
\hat{E}_{\mathrm{in}}^{(-)}&(x,y,z,t)=\sqrt{\frac{\pi\hbar\omega_0}{c}}\left\{U_{00}(x,y,z,t)A\right.\\
&\left.+U'_{10}(x,y,z,t;Z)
\left[\hat{a}^{10}_1(z,t;Z)-\mathrm{i}\hat{a}^{10}_2(z,t;Z)\right]\right\}.\nonumber
\end{align}
It can be easily shown that $\hat{a}^{lm}_j(\Omega;Z)\,\,(j=1,2)$, which is defined by the Fourier transform $\hat{a}^{lm}_j(z,t;Z)=\int_0^\infty\frac{\mathrm{d}\Omega}{2\pi}\left(\hat{a}^{lm}_j(\Omega;Z)\mathrm{e}^{\mathrm{i}\Omega(z/c-t)}+\mathrm{H.c.}\right)$,
satisfies the same commutation relations as Eq. (\ref{com}). Then, from Eq. (\ref{10d}), which is a classical expression for the transverse displacement and tilt, quantum operators for the transverse displacement $\hat{\delta x}$ and tilt $\hat{\delta\theta}$ at $z=Z$ can be identified as 
\begin{align}
\frac{\hat{\delta x}(t,Z)}{w(Z)}=\frac{\hat{a}^{10}_1(Z,t;Z)}{A},\qquad
\frac{\hat{\delta\theta}(t,Z)}{\alpha(Z)}=-\frac{\hat{a}^{10}_2(Z,t;Z)}{A},
\end{align}
respectively. 
The Fourier transforms of $\hat{\delta x}$ and $\hat{\delta\theta}$ are
\begin{align}
\frac{\hat{\delta x}(\Omega,Z)}{w(Z)}=\frac{\hat{a}^{10}_1(\Omega;Z)}{A},\qquad
\frac{\hat{\delta \theta}(\Omega,Z)}{\alpha(Z)}=-\frac{\hat{a}^{10}_2(\Omega;Z)}{A}.\label{disp}
\end{align}

\begin{table*}
\caption{Analogy between the quantum noise in the measurement of the translational motion and the angular motion of a mirror. This table summarizes HG modes that contribute to the quantum noise, fluctuations caused by HG mode vacuum fields, expressions of ponderomotive squeezing, and parameters of homodyne detection, for the two types of measurement.\label{table}}
\begin{ruledtabular}
\begin{tabular}{ccccc}
 HG mode of vacuum&fluctuation: $\hat{a}_1$&fluctuation: $\hat{a}_2$&ponderomotive squeezing &homodyne detection\\ \hline
 00 mode&amplitude&phase&$\hat{a}^{00}_2\rightarrow-\kappa_0\hat{a}^{00}_1+\hat{a}^{00}_2$&control of homodyne angle $\eta$\\
 10 mode&transverse displacement
 &transverse tilt&$\hat{a}^{10}_2\rightarrow-\kappa_1\hat{a}^{10}_1+\hat{a}^{10}_2$&choice of gouy phase shift $\psi$\\
\end{tabular}
\end{ruledtabular}
\end{table*}

Next, we consider the angular motion of the mirror located at $z=-Z_1$ and the reflected laser light, as shown in Fig. \ref{refl2}. With Eq. (\ref{disp}), we can calculate torque on the mirror caused by radiation pressure of the laser light, and the equation of angular motion in frequency domain turns out to be
\begin{align}
\hat{\theta}_\mathrm{mir}(\Omega)
&=\hat{\theta}_\mathrm{ex}(\Omega)+\frac{\hbar\omega_0Aw(Z_1)}{cI\Omega^2}\hat{a}^{10}_1(\Omega;-Z_1).
\end{align} 
Here $\hat{\theta}_\mathrm{mir}(\Omega)$ is the angular tilt of the mirror, $\hat{\theta}_\mathrm{ex}(\Omega)$ is the angular tilt of the mirror caused by an external force other than radiation pressure, $I$ is moment of inertia of the mirror, and the fact that radiation pressure force is $\hbar\omega_0A^2/c$ is used. Then, since the transverse displacement and tilt of the light reflected by the mirror are $\hat{\delta x}$ and $\hat{\delta\theta}+2\hat{\theta}_\mathrm{mir}$, respectively, we can express the electric field of the reflected light using Eq. (\ref{10d}) as
\begin{align}
\hat{E}_{\mathrm{out}}^{(-)}(x,y,z,t)&=\sqrt{\frac{\pi\hbar\omega_0}{c}}A\biggl\{U_{00}(x,y,z,t)\nonumber\\
+U'_{10}(x,y,z,t;&Z_1)\left[\frac{\hat{\delta x}(-Z_1)}{w(Z_1)}+\mathrm{i}\frac{\hat{\delta\theta}(-Z_1)+2\hat{\theta}_\mathrm{mir}}{\alpha(Z_1)}\right]\biggr\}\label{out}\\
&\equiv\sqrt{\frac{\pi\hbar\omega_0}{c}}\biggl\{U_{00}(x,y,z,t)A\nonumber\\
+U'_{10}(x,y,z,t;&Z_1)\left[\hat{b}^{10}_1(z,t;Z_1)-\mathrm{i}\hat{b}^{10}_2(z,t;Z_1)\right]\biggr\}.
\end{align}
Defining $\hat{b}^{10}_j(\Omega;Z_1)\,\,(j=1,2)$ by the Fourier transform $\hat{b}^{10}_j(z,t;Z_1)=\int_0^\infty\frac{\mathrm{d}\Omega}{2\pi}\left(\hat{b}^{10}_j(\Omega;Z_1)\mathrm{e}^{\mathrm{i}\Omega(z/c-t)}+\mathrm{H.c.}\right),$
we obtain the following equations similar to Eqs. (\ref{pdr1}) and (\ref{pdr2}):
\begin{align}
\hat{b}^{10}_1(\Omega;Z_1)&=\hat{a}^{10}_1(\Omega;-Z_1),\\
\hat{b}^{10}_2(\Omega;Z_1)&=-\kappa_1\hat{a}^{10}_1(\Omega;-Z_1)+\hat{a}^{10}_2(\Omega;-Z_1)\nonumber\\
&+\sqrt{2\kappa_1}\frac{\hat{\theta}_\mathrm{ex}(\Omega)}{\theta_\mathrm{SQL}},
\end{align}
where 
$\kappa_1=4I_0w(Z_1)/Ic\Omega^2\alpha(Z_1)$, and $\theta_\mathrm{SQL}=\sqrt{2\hbar/I\Omega^2}$.
Since the term $-\kappa_1\hat{a}^{10}_1(\Omega;-Z_1)$ exists in the expression of $\hat{b}^{10}_2(\Omega;Z_1)$ in the same way as Eq. (\ref{pdr2}), the 10 mode vacuum field of the light reflected by the mirror can be regarded as ponderomotively squeezed.

Then, we consider measurement of $\hat{\theta}_\mathrm{ex}$ by detecting the transverse displacement of the reflected light at $z=Z_\mathrm{meas}$ and the expression of quantum noise of this measurement. To obtain the transverse displacement of the reflected light at $z=Z_\mathrm{meas}$, we write $\hat{E}_{\mathrm{out}}^{(-)}$ as
\begin{align}
&\hat{E}_{\mathrm{out}}^{(-)}(x,y,z,t)\nonumber\\
&=\sqrt{\frac{\pi\hbar\omega_0}{c}}A\Biggl\{U_{00}(x,y,z,t)+U'_{10}(x,y,z,t;Z_\mathrm{meas})\nonumber\\
&\times\left[\frac{\hat{x}_\mathrm{meas}}{w(Z_\mathrm{meas})}+\mathrm{i}\frac{\hat{\theta}_\mathrm{meas}}{\alpha(Z_\mathrm{meas})}\right]\Biggr\}\label{this}.
\end{align}
According to the interpretation of Eq. (\ref{10d}), $\hat{x}_\mathrm{meas}$ and $\hat{\theta}_\mathrm{meas}$ are transverse displacement and tilt at $z=Z_\mathrm{meas}$, respectively. Comparing Eq. (\ref{this}) with Eq. (\ref{out}), we obtain
\begin{align}
\hat{x}_\mathrm{meas}(\Omega)&=-\frac{2w(Z_\mathrm{meas})\sin\psi}{\alpha(Z_1)}\nonumber\\ \times\biggl(-&\frac{\alpha(Z_1)\hat{\delta x}(-Z_1)}{2w(Z_1)}\cot\psi+\frac{\hat{\delta\theta}(-Z_1)}{2}+\hat{\theta}_\mathrm{mir}\biggr)\\
&=-\frac{2w(Z_\mathrm{meas})\sin\psi}{\alpha(Z_1)}\biggl(\hat{\theta}_\mathrm{ex}(\Omega)\nonumber\\-\frac{\alpha(Z_1)}{2A}&[(-\kappa_1+\cot\psi)\hat{a}^{10}_1(\Omega;-Z_1)+\hat{a}^{10}_2(\Omega;-Z_1)]\biggr),
\end{align}
where $\psi\equiv\zeta(Z_\mathrm{meas})-\zeta(Z_1)$.
Using Eq. (\ref{spectrum}) we obtain the noise spectral density of the measurement of $\hat{\theta}_\mathrm{ex}$ as
\begin{equation}
S_{\theta,\psi}(\Omega)=\frac{\theta^2_\mathrm{SQL}}{2\kappa_1}[(-\kappa_1+\cot\psi)^2+1].\label{10noise}
\end{equation}
Note that Eq. (\ref{10noise}) takes exactly the same form as Eq. (\ref{00noise}). Table \ref{table} summarizes the analogy between the quantum noise in the measurement of the translational motion and the angular motion of a mirror. 
This expression for the noise spectrum shows that $\psi$, the Gouy phase shift from the mirror to the position of detection, corresponds to the homodyne angle of the measurement of the translational motion of a mirror. Therefore, measurement of the transverse displacement of the light reflected by the mirror at modest distance from the mirror, i.e. $\psi\ne\pi/2$, can be regarded as homodyne detection.
If $\psi=\pi/2$, which corresponds, for example, to the case that the separation between the mirror and the position of detection is sufficiently larger than the Rayleigh range and the light is detected at the beam waist, i.e. $\zeta(Z_\mathrm{meas})\simeq0$, $\zeta(Z_1)\simeq -\pi/2$, Eq. (\ref{10noise}) is simplified as
\begin{equation}
S_{\theta,\pi/2}(\Omega)=\frac{\theta^2_\mathrm{SQL}}{2\kappa_1}(\kappa_1^2+1)\ge\theta^2_\mathrm{SQL}.\label{sql}
\end{equation}
The first and the second terms inside the brackets originate in the fluctuation of the angle of the mirror caused by the fluctuation of the transverse displacement of the incident light and the fluctuation of the angle that the incident light has had before hitting the mirror, respectively. Therefore, they are analogs of radiation pressure noise and shot noise, respectively. We call the first term radiation torque noise. From Eq. (\ref{sql}) We can see that $S_{\theta,\pi/2}(\Omega)$ is always larger than $\theta^2_\mathrm{SQL}$ even if the power or the beam width of the incident laser light is varied. In other words, this formula shows that $\theta_\mathrm{SQL}$ is the SQL of the measurement of the angle of a mirror. 
Equation (\ref{10noise}) also indicates that at the frequency where the condition $-\kappa_1+\cot\psi=0$ holds, radiation torque noise is completely cancelled. Therefore, by choosing the appropriate position of measurement $Z_\mathrm{meas}$, radiation torque noise at a certain frequency can be cancelled and the SQL can be beaten, in a similar way to the radiation pressure noise cancellation by choosing the appropriate homodyne angle.

We will also discuss these results in terms of interaction between orbital angular momentum of light and motion of the mirror reflecting the light. Backaction to the suspended mirror in our configuration can be explained by interaction between the transverse $(y)$ component of optical angular momentum of the laser light and angular motion of the mirror. In fact, it was shown that light which is characterized by a linear combination of HG00 mode and 10 mode has a non-zero $y$ component of orbital angular momentum \cite{Aiello2010}. In contrast, interaction between the longitudinal ($z$) component of orbital angular momentum of light and a mirror made of a spiral phase element that reflects the light has been investigated to utilize it to cool the rotational motion of the mirror about the $z$-axis, in which direction the light propagates \cite{Bhattacharya2007}. Therefore, comparison of our results and this previous work indicates that our present approach might also be applicable to addressing the quantum limit in the measurement of rotation of a mirror about the $z$-axis.


\begin{acknowledgments}
We thank K. Komori, M. Nakano, T. Akutsu, and K. Somiya for fruitful discussions and useful suggestions, and K. Craig for English proofreading. This work was supported by MEXT, JSPS Leading-edge Research Infrastructure Program, JSPS Grant-in-Aid for Specially Promoted Research 26000005, MEXT Grant-in-Aid for Scientific Research on Innovative Areas 24103005, JSPS Core-to-Core Program, A. Advanced Research Networks, JSPS KAKENHI Grant Number 23340077 and 26287038, and the joint research program of the Institute for Cosmic Ray Research, University of Tokyo.

\end{acknowledgments}


\begin{thebibliography}{9}%
\makeatletter
\providecommand \@ifxundefined [1]{%
 \@ifx{#1\undefined}
}%
\providecommand \@ifnum [1]{%
 \ifnum #1\expandafter \@firstoftwo
 \else \expandafter \@secondoftwo
 \fi
}%
\providecommand \@ifx [1]{%
 \ifx #1\expandafter \@firstoftwo
 \else \expandafter \@secondoftwo
 \fi
}%
\providecommand \natexlab [1]{#1}%
\providecommand \enquote  [1]{``#1''}%
\providecommand \bibnamefont  [1]{#1}%
\providecommand \bibfnamefont [1]{#1}%
\providecommand \citenamefont [1]{#1}%
\providecommand \href@noop [0]{\@secondoftwo}%
\providecommand \href [0]{\begingroup \@sanitize@url \@href}%
\providecommand \@href[1]{\@@startlink{#1}\@@href}%
\providecommand \@@href[1]{\endgroup#1\@@endlink}%
\providecommand \@sanitize@url [0]{\catcode `\\12\catcode `\$12\catcode
  `\&12\catcode `\#12\catcode `\^12\catcode `\_12\catcode `\%12\relax}%
\providecommand \@@startlink[1]{}%
\providecommand \@@endlink[0]{}%
\providecommand \url  [0]{\begingroup\@sanitize@url \@url }%
\providecommand \@url [1]{\endgroup\@href {#1}{\urlprefix }}%
\providecommand \urlprefix  [0]{URL }%
\providecommand \Eprint [0]{\href }%
\providecommand \doibase [0]{http://dx.doi.org/}%
\providecommand \selectlanguage [0]{\@gobble}%
\providecommand \bibinfo  [0]{\@secondoftwo}%
\providecommand \bibfield  [0]{\@secondoftwo}%
\providecommand \translation [1]{[#1]}%
\providecommand \BibitemOpen [0]{}%
\providecommand \bibitemStop [0]{}%
\providecommand \bibitemNoStop [0]{.\EOS\space}%
\providecommand \EOS [0]{\spacefactor3000\relax}%
\providecommand \BibitemShut  [1]{\csname bibitem#1\endcsname}%
\let\auto@bib@innerbib\@empty
\bibitem [{\citenamefont {{V. B. Braginsky , F. Y.
  Khalili}}(1995)}]{Braginsky1995}%
  \BibitemOpen
  \bibfield  {author} {\bibinfo {author} {
  \bibnamefont {{V. B. Braginsky, F. Y. Khalili, edited by K. S. Thorne}}},\ }\href@noop {}
   {\bibinfo {title} {{Quantum Measurement}}} (\bibinfo  {publisher}
  {Cambridge},\ \bibinfo {year} {1995})\BibitemShut {NoStop}%
\bibitem [{\citenamefont {Danilishin}\ and\ \citenamefont
  {Khalili}(2012)}]{Danilishin2012}%
  \BibitemOpen
  \bibfield  {author} {\bibinfo {author} {\bibfnamefont {S.~L.}\ \bibnamefont
  {Danilishin}}\ and\ \bibinfo {author} {\bibfnamefont {F.~Y.}\ \bibnamefont
  {Khalili}},\ }\href {\doibase 10.12942/lrr-2012-5} {\bibfield  {journal}
  {\bibinfo  {journal} {Living Reviews in Relativity}\ }\textbf {\bibinfo
  {volume} {15}},\ \bibinfo {pages} {1} (\bibinfo {year} {2012})},\ \Eprint
  {http://arxiv.org/abs/1203.1706} {arXiv:1203.1706} \BibitemShut {NoStop}%
\bibitem [{\citenamefont {Chen}(2013)}]{Chen2013a}%
  \BibitemOpen
  \bibfield  {author} {\bibinfo {author} {\bibfnamefont {Y.}~\bibnamefont
  {Chen}},\ }\href {\doibase 10.1088/0953-4075/46/10/104001} {\bibfield
  {journal} {\bibinfo  {journal} {Journal of Physics B: Atomic, Molecular and
  Optical Physics}\ }\textbf {\bibinfo {volume} {46}},\ \bibinfo {pages}
  {104001} (\bibinfo {year} {2013})},\ \Eprint {http://arxiv.org/abs/1302.1924}
  {arXiv:1302.1924} \BibitemShut {NoStop}%
\bibitem [{\citenamefont {Kimble}\ \emph {et~al.}(2002)\citenamefont {Kimble},
  \citenamefont {Levin}, \citenamefont {Matsko}, \citenamefont {Thorne},\ and\
  \citenamefont {Vyatchanin}}]{Kimble2002}%
  \BibitemOpen
  \bibfield  {author} {\bibinfo {author} {\bibfnamefont {H.~J.}\ \bibnamefont
  {Kimble}}, \bibinfo {author} {\bibfnamefont {Y.}~\bibnamefont {Levin}},
  \bibinfo {author} {\bibfnamefont {A.~B.}\ \bibnamefont {Matsko}}, \bibinfo
  {author} {\bibfnamefont {K.~S.}\ \bibnamefont {Thorne}}, \ and\ \bibinfo
  {author} {\bibfnamefont {S.~P.}\ \bibnamefont {Vyatchanin}},\ }\href
  {\doibase 10.1103/PhysRevD.65.022002} {\bibfield  {journal} {\bibinfo
  {journal} {Physical Review D}\ }\textbf {\bibinfo {volume} {65}},\ \bibinfo
  {pages} {022002} (\bibinfo {year} {2001})},\ \Eprint
  {http://arxiv.org/abs/0008026} {arXiv:0008026 [gr-qc]} \BibitemShut {NoStop}%
\bibitem [{\citenamefont {Kogelnik}\ and\ \citenamefont
  {Li}(1966)}]{Kogelnik1966}%
  \BibitemOpen
  \bibfield  {author} {\bibinfo {author} {\bibfnamefont {H.}~\bibnamefont
  {Kogelnik}}\ and\ \bibinfo {author} {\bibfnamefont {T.}~\bibnamefont {Li}},\
  }\href {\doibase 10.1364/AO.5.001550} {\bibfield  {journal} {\bibinfo
  {journal} {Applied optics}\ }\textbf {\bibinfo {volume} {5}},\ \bibinfo
  {pages} {1550} (\bibinfo {year} {1966})}\BibitemShut {NoStop}%
\bibitem [{\citenamefont {Morrison}\ \emph {et~al.}(1994)\citenamefont
  {Morrison}, \citenamefont {Meers}, \citenamefont {Robertson},\ and\
  \citenamefont {Ward}}]{Morrison1994}%
  \BibitemOpen
  \bibfield  {author} {\bibinfo {author} {\bibfnamefont {E.}~\bibnamefont
  {Morrison}}, \bibinfo {author} {\bibfnamefont {B.~J.}\ \bibnamefont {Meers}},
  \bibinfo {author} {\bibfnamefont {D.~I.}\ \bibnamefont {Robertson}}, \ and\
  \bibinfo {author} {\bibfnamefont {H.}~\bibnamefont {Ward}},\ }\href {\doibase
  10.1364/AO.33.005041} {\bibfield  {journal} {\bibinfo  {journal} {Applied
  optics}\ }\textbf {\bibinfo {volume} {33}},\ \bibinfo {pages} {5041}
  (\bibinfo {year} {1994})}\BibitemShut {NoStop}%
\bibitem [{\citenamefont {W{\"{u}}nsche}(2004)}]{Wunsche2004}%
  \BibitemOpen
  \bibfield  {author} {\bibinfo {author} {\bibfnamefont {A.}~\bibnamefont
  {W{\"{u}}nsche}},\ }\href {\doibase 10.1088/1464-4266/6/3/009} {\bibfield
  {journal} {\bibinfo  {journal} {Journal of Optics B: Quantum and
  Semiclassical Optics}\ }\textbf {\bibinfo {volume} {6}},\ \bibinfo {pages}
  {S47} (\bibinfo {year} {2004})}\BibitemShut {NoStop}%
\bibitem [{\citenamefont {Aiello}\ \emph {et~al.}(2010)\citenamefont {Aiello},
  \citenamefont {Marquardt},\ and\ \citenamefont {Leuchs}}]{Aiello2010}%
  \BibitemOpen
  \bibfield  {author} {\bibinfo {author} {\bibfnamefont {A.}~\bibnamefont
  {Aiello}}, \bibinfo {author} {\bibfnamefont {C.}~\bibnamefont {Marquardt}}, \
  and\ \bibinfo {author} {\bibfnamefont {G.}~\bibnamefont {Leuchs}},\ }\href
  {\doibase 10.1103/PhysRevA.81.053838} {\bibfield  {journal} {\bibinfo
  {journal} {Physical Review A}\ }\textbf {\bibinfo {volume} {81}},\ \bibinfo
  {pages} {053838} (\bibinfo {year} {2010})},\ \Eprint
  {http://arxiv.org/abs/1003.0989} {arXiv:1003.0989} \BibitemShut {NoStop}%
\bibitem [{\citenamefont {Miao}(2010)}]{Miao2010}%
  \BibitemOpen
  \bibfield  {author} {\bibinfo {author} {\bibfnamefont {H.}~\bibnamefont
  {Miao}},\ } {\bibinfo {title} {{Exploring Macroscopic Quantum Mechanics
  in Optomechanical Devices}}}, \href@noop {} {Springer Thesis},\  (\bibinfo {year}
  {Springer, 2012})\BibitemShut {NoStop}%
\bibitem [{\citenamefont {Bhattacharya}\ and\ \citenamefont {Meystre}(2007)}]{Bhattacharya2007}%
  \BibitemOpen
  \bibfield  {author} {\bibinfo {author} {\bibfnamefont {M.}~\bibnamefont
  {Bhattacharya}}\
  and\ \bibinfo {author} {\bibfnamefont {P.}~\bibnamefont {Meystre}},\ }\href
  {\doibase 10.1103/PhysRevLett.99.153603} {\bibfield  {journal} {\bibinfo
  {journal} {Physical Review Letters}\ }\textbf {\bibinfo {volume} {99}},\ \bibinfo
  {pages} {153603} (\bibinfo {year} {2007})},\ \Eprint
  {http://arxiv.org/abs/0705.0784} {arXiv:0705.0784} \BibitemShut {NoStop}%
\end{thebibliography}
\end{document}